\documentclass[%
	reprint,
	superscriptaddress,
	amsmath,amssymb,
	aps,
	prd,
]{revtex4-1}

\usepackage{gensymb}
\usepackage{graphicx}
\usepackage{multirow}
\usepackage{dcolumn}
\usepackage{bm}
\usepackage[mathlines]{lineno}

\begin{document}

\preprint{APS/123-QED}

\title{Measurement of neutrino-oxygen neutral-current quasi-elastic cross section using atmospheric neutrinos at Super-Kamiokande}

\newcommand{\AFFicrr}{\affiliation{Kamioka Observatory, Institute for Cosmic Ray Research, University of Tokyo, Kamioka, Gifu 506-1205, Japan}}
\newcommand{\AFFkashiwa}{\affiliation{Research Center for Cosmic Neutrinos, Institute for Cosmic Ray Research, University of Tokyo, Kashiwa, Chiba 277-8582, Japan}}
\newcommand{\AFFipmu}{\affiliation{Kavli Institute for the Physics and
Mathematics of the Universe (WPI), The University of Tokyo Institutes for Advanced Study,
University of Tokyo, Kashiwa, Chiba 277-8583, Japan }}
\newcommand{\AFFmad}{\affiliation{Department of Theoretical Physics, University Autonoma Madrid, 28049 Madrid, Spain}}
\newcommand{\AFFubc}{\affiliation{Department of Physics and Astronomy, University of British Columbia, Vancouver, BC, V6T1Z4, Canada}}
\newcommand{\AFFbu}{\affiliation{Department of Physics, Boston University, Boston, MA 02215, USA}}
\newcommand{\AFFuci}{\affiliation{Department of Physics and Astronomy, University of California, Irvine, Irvine, CA 92697-4575, USA }}
\newcommand{\AFFcsu}{\affiliation{Department of Physics, California State University, Dominguez Hills, Carson, CA 90747, USA}}
\newcommand{\AFFcnm}{\affiliation{Department of Physics, Chonnam National University, Kwangju 500-757, Korea}}
\newcommand{\AFFduke}{\affiliation{Department of Physics, Duke University, Durham NC 27708, USA}}
\newcommand{\AFFfukuoka}{\affiliation{Junior College, Fukuoka Institute of Technology, Fukuoka, Fukuoka 811-0295, Japan}}
\newcommand{\AFFgifu}{\affiliation{Department of Physics, Gifu University, Gifu, Gifu 501-1193, Japan}}
\newcommand{\AFFgist}{\affiliation{GIST College, Gwangju Institute of Science and Technology, Gwangju 500-712, Korea}}
\newcommand{\AFFuh}{\affiliation{Department of Physics and Astronomy, University of Hawaii, Honolulu, HI 96822, USA}}
\newcommand{\AFFicl}{\affiliation{Department of Physics, Imperial College London , London, SW7 2AZ, United Kingdom }}
\newcommand{\AFFkek}{\affiliation{High Energy Accelerator Research Organization (KEK), Tsukuba, Ibaraki 305-0801, Japan }}
\newcommand{\AFFkobe}{\affiliation{Department of Physics, Kobe University, Kobe, Hyogo 657-8501, Japan}}
\newcommand{\AFFkyoto}{\affiliation{Department of Physics, Kyoto University, Kyoto, Kyoto 606-8502, Japan}}
\newcommand{\AFFliv}{\affiliation{Department of Physics, University of Liverpool, Liverpool, L69 7ZE, United Kingdom}}
\newcommand{\AFFmiyagi}{\affiliation{Department of Physics, Miyagi University of Education, Sendai, Miyagi 980-0845, Japan}}
\newcommand{\AFFnagoya}{\affiliation{Institute for Space-Earth Environmental Research, Nagoya University, Nagoya, Aichi 464-8602, Japan}}
\newcommand{\AFFkmi}{\affiliation{Kobayashi-Maskawa Institute for the Origin of Particles and the Universe, Nagoya University, Nagoya, Aichi 464-8602, Japan}}
\newcommand{\AFFpol}{\affiliation{National Centre For Nuclear Research, 00-681 Warsaw, Poland}}
\newcommand{\AFFsuny}{\affiliation{Department of Physics and Astronomy, State University of New York at Stony Brook, NY 11794-3800, USA}}
\newcommand{\AFFokayama}{\affiliation{Department of Physics, Okayama University, Okayama, Okayama 700-8530, Japan }}
\newcommand{\AFFosaka}{\affiliation{Department of Physics, Osaka University, Toyonaka, Osaka 560-0043, Japan}}
\newcommand{\AFFox}{\affiliation{Department of Physics, Oxford University, Oxford, OX1 3PU, United Kingdom}}
\newcommand{\AFFqmul}{\affiliation{School of Physics and Astronomy, Queen Mary University of London, London, E1 4NS, United Kingdom}}
\newcommand{\AFFregina}{\affiliation{Department of Physics, University of Regina, 3737 Wascana Parkway, Regina, SK, S4SOA2, Canada}}
\newcommand{\AFFseoul}{\affiliation{Department of Physics, Seoul National University, Seoul 151-742, Korea}}
\newcommand{\AFFsheff}{\affiliation{Department of Physics and Astronomy, University of Sheffield, S3 7RH, Sheffield, United Kingdom}}
\newcommand{\AFFshizuokasc}{\affiliation{Department of Informatics in
Social Welfare, Shizuoka University of Welfare, Yaizu, Shizuoka, 425-8611, Japan}}
\newcommand{\AFFstfc}{\affiliation{STFC, Rutherford Appleton Laboratory, Harwell Oxford, and Daresbury Laboratory, Warrington, OX11 0QX, United Kingdom}}
\newcommand{\AFFskk}{\affiliation{Department of Physics, Sungkyunkwan University, Suwon 440-746, Korea}}
\newcommand{\AFFtokyo}{\affiliation{The University of Tokyo, Bunkyo, Tokyo 113-0033, Japan }}
\newcommand{\AFFtodai}{\affiliation{Department of Physics, University of Tokyo, Bunkyo, Tokyo 113-0033, Japan }}
\newcommand{\AFFtit}{\affiliation{Department of Physics,Tokyo Institute of Technology, Meguro, Tokyo 152-8551, Japan }}
\newcommand{\AFFtus}{\affiliation{Department of Physics, Faculty of Science and Technology, Tokyo University of Science, Noda, Chiba 278-8510, Japan }}
\newcommand{\AFFtoronto}{\affiliation{Department of Physics, University of Toronto, ON, M5S 1A7, Canada }}
\newcommand{\AFFtriumf}{\affiliation{TRIUMF, 4004 Wesbrook Mall, Vancouver, BC, V6T2A3, Canada }}
\newcommand{\AFFtokai}{\affiliation{Department of Physics, Tokai University, Hiratsuka, Kanagawa 259-1292, Japan}}
\newcommand{\AFFtsinghua}{\affiliation{Department of Engineering Physics, Tsinghua University, Beijing, 100084, China}}
\newcommand{\AFFynu}{\affiliation{Faculty of Engineering, Yokohama National University, Yokohama, 240-8501, Japan}}
\newcommand{\AFFllr}{\affiliation{Ecole Polytechnique, IN2P3-CNRS, Laboratoire Leprince-Ringuet, F-91120 Palaiseau, France }}
\newcommand{\AFFbari}{\affiliation{ Dipartimento Interuniversitario di Fisica, INFN Sezione di Bari and Universit\`a e Politecnico di Bari, I-70125, Bari, Italy}}
\newcommand{\AFFnapoli}{\affiliation{Dipartimento di Fisica, INFN Sezione di Napoli and Universit\`a di Napoli, I-80126, Napoli, Italy}}
\newcommand{\AFFroma}{\affiliation{INFN Sezione di Roma and Universit\`a di Roma ``La Sapienza'', I-00185, Roma, Italy}}
\newcommand{\AFFpadova}{\affiliation{Dipartimento di Fisica, INFN Sezione di Padova and Universit\`a di Padova, I-35131, Padova, Italy}}

\AFFicrr
\AFFkashiwa
\AFFmad
\AFFbu
\AFFubc
\AFFuci
\AFFcsu
\AFFcnm
\AFFduke
\AFFllr
\AFFfukuoka
\AFFgifu
\AFFgist
\AFFuh
\AFFicl
\AFFbari
\AFFnapoli
\AFFpadova
\AFFroma
\AFFkek
\AFFkobe
\AFFkyoto
\AFFliv
\AFFmiyagi
\AFFnagoya
\AFFkmi
\AFFpol
\AFFsuny
\AFFokayama
\AFFosaka
\AFFox
\AFFqmul
\AFFregina
\AFFseoul
\AFFsheff
\AFFshizuokasc
\AFFstfc
\AFFskk
\AFFtokai
\AFFtokyo
\AFFtodai
\AFFipmu
\AFFtit
\AFFtus
\AFFtoronto
\AFFtriumf
\AFFtsinghua
\AFFynu

\author{L.~Wan}
\AFFtsinghua
\author{K.~Abe}
\AFFicrr
\AFFipmu
\author{C.~Bronner}
\AFFicrr
\author{Y.~Hayato}
\AFFicrr
\AFFipmu
\author{M.~Ikeda}
\AFFicrr
\author{K.~Iyogi}
\AFFicrr 
\author{J.~Kameda}
\AFFicrr
\AFFipmu 
\author{Y.~Kato}
\AFFicrr
\author{Y.~Kishimoto}
\AFFicrr
\AFFipmu 
\author{Ll.~Marti}
\AFFicrr
% ...


\collaboration{The Super-Kamiokande Collaboration}
\noaffiliation

\date{\today}

\begin{abstract}
	Neutral current (NC) interactions of atmospheric neutrinos on oxygen form one of the major backgrounds in the search for supernova relic neutrinos with water-based Cherenkov detectors.
	The NC channel is dominated by neutrino quasi-elastic (NCQE) scattering off nucleons inside $^{16}$O nuclei.
	In this paper we report the first measurement of NCQE cross section using atmospheric neutrinos at Super-Kamiokande (SK).
	The measurement used 2,778 live days of SK-IV data with a fiducial volume of 22.5 kiloton water.
	Within the visible energy window of 7.5-29.5 MeV, we observed $117$ events compared to the expected $71.9$ NCQE signal and $53.1$ background events.
	Weighted by the atmospheric neutrino spectrum from 160 MeV to 10 GeV, the flux averaged NCQE cross section is measured to be $(1.01\pm0.17(\text{stat.})^{+0.78}_{-0.30}(\text{sys.}))\times10^{-38}$ cm$^2$.
	\begin{description}
		\item[DOI]
	\end{description}
\end{abstract}
\maketitle

\section{Introduction}

Neutral current quasi-elastic interaction (NCQE) of atmospheric neutrinos with $^{16}$O is one of the major interaction channels in water based neutrino detectors for neutrinos with several hundred MeV energy~\cite{Ankowski:2011ei}.
The interaction processes can be written as 
\begin{equation}
	\begin{aligned}
		\nu+^{16}\text{O}&\to\nu+^{15}\text{O}+n+\gamma,\\
		\nu+^{16}\text{O}&\to\nu+^{15}\text{N}+p+\gamma,
	\end{aligned}
\end{equation}
in which neutrinos knock out nucleons from oxygen and the residual nuclei are likely to produce de-excitation $\gamma$'s.
The $\gamma$ ray propagates in water and is detected by the Cherenkov light of the electrons or positrons from compton scattering and pair production.
The emitted proton is below Cherenkov threshold, while the emitted neutron will be captured on hydrogen, releasing a 2.2 MeV $\gamma$-ray.
We use these $\gamma$-rays to tag the interaction.

Supernova relic neutrinos (SRN), also known as diffused supernova neutrino background (DSNB), are neutrinos emitted from all past core-collapse supernovae~\cite{Ando:2004hc}.
The detection of SRNs via inverse beta decay (IBD, $\bar\nu_e+p\to n+e^+$) is a goal of current and future large neutrino detectors~\cite{Horiuchi:2008jz}.
In water Cherenkov detectors such as Super-Kamiokande (SK), where $e/\gamma$ discrimination is feasible but still challenging, NCQE interactions of atmospheric neutrinos with oxygen form a significant background to SRN searches as well as other rare signal searches~\cite{Kocharov:1991cn, Abe:2018mic, Abe:2016jwn, Cui:2017ytb}.

Theoretical calculations of the NCQE cross section on oxygen exist for both low~\cite{Langanke:1995he, Beacom:1998ya, Kolbe:2002gk} and high~\cite{Ankowski:2011ei} energies and a measurement using the T2K beam~\cite{Abe:2014dyd} is in good agreement with the predictions.
However, no previous measurement of this process has been performed using the atmospheric neutrino flux, where NCQE interactions are of relevance for SRN searches.
In this paper, we report the first measurement of the NCQE cross section on oxygen using atmospheric neutrinos. 
This measurement is particularly important for SRN searches in future water Cherenkov experiments such as SK-Gd~\cite{Beacom:2003nk, Sekiya:2017lgj} and Hyper-Kamiokande~\cite{Abe:2011ts}.

The paper is organized as follows:
Section~\ref{sec:SK} introduces the Super-Kamiokande experiment.
Section~\ref{sec:MC} illustrates the simulation of NCQE events, the Monte-Carlo (MC) setup, and the corresponding features expected.
The NCQE sample is extracted from data and compared with MC in Section~\ref{sec:data}.
Section~\ref{sec:analysis} presents the measurement of NCQE cross section, and discusses applications and future improvements.
Section~\ref{sec:conclusion} concludes the paper.

\section{The Super-Kamiokande experiment}
\label{sec:SK}

Super-Kamiokande is a cylindrical 50-kiloton water Cherenkov detector located in Kamioka, Japan, shielded by 2,700 meter water equivalent overburden~\cite{Fukuda:2002uc}.
SK consists of an outer detector (OD) instrumented with 1885 8-inch PMTs, optically separated from the inner detector (ID), which is viewed by 11,129 20-inch PMTs.

SK started data taking in 1996, and since then has undergone four data-taking phases: SK-I, II, III and IV.
This measurement uses data in SK-IV only, collected from October 2008 to October 2017.
SK-IV started in the summer of 2008, when new front-end electronics and data processing system were installed.
The data acquisition time window for a typical event above 7.5 MeV in SK-IV is $[-5, 35]\ \mu$s from the trigger timing~\cite{5446533}.
If the kinetic energy of an event is above 9.5 MeV (or 7.5 MeV after the summer of 2011), and if the event is not a cosmic-ray muon, a special high energy trigger (SHE) and a following after-trigger with 500 $\mu$s time window are issued, allowing for the detection of delayed-coincidence 2.2 MeV $\gamma$ signals from neutron captures on hydrogen within a 535 $\mu$s search window.

The interaction vertex of a low-energy event is reconstructed with a time-of-flight (TOF) based algorithm~\cite{Smy:2007maa}, with a vertex resolution of 65~cm for a 7.5 MeV electron, improving as the energy increases. 
The energy is reconstructed from the number of detected Cherenkov photons corrected for water attenuation, photo-coverage, and PMT response.
A detailed description of reconstruction and calibration for low-energy events can be found in Ref.~\cite{Abe:2016nxk}.

\section{Simulation}
\label{sec:MC}

The MC simulation of atmospheric neutrino events is performed in two stages.
First, an event generator models the interactions of atmospheric neutrinos in water.
Final-state particles resulting from these interactions are then tracked through a simulation of the ID to model the detector response.
This analysis uses a MC sample equivalent to 500 years of live-time with the SK-IV setup.

\subsection{Atmospheric neutrino flux}

Super-Kamiokande performed a comprehensive study of the atmospheric neutrino flux in the energy region from sub-GeV up to several TeV at the Kamioka Observatory~\cite{Richard:2015aua}.
The measured observables, including event rate, energy spectrum, and directionality, are consistent with the theoretical prediction from the HKKM model~\cite{Honda:2006qj, Honda:2011nf}, after accounting for neutrino oscillation.
Uncertainties on the flux are taken from the SK measurement and the neutrino/antineutrino ratio uncertainty is derived from the theoretical prediction.

\subsection{Neutrino interactions}

Atmospheric neutrino interactions with constituents of water molecules in SK are simulated by the NEUT generator~\cite{Hayato:2009zz}, with several modifications in the neutral current model~\cite{Ueno:2012}.
We describe here the simulation for NCQE.

\begin{figure}[htbp] 
	\centering
	\includegraphics[width=0.89\columnwidth]{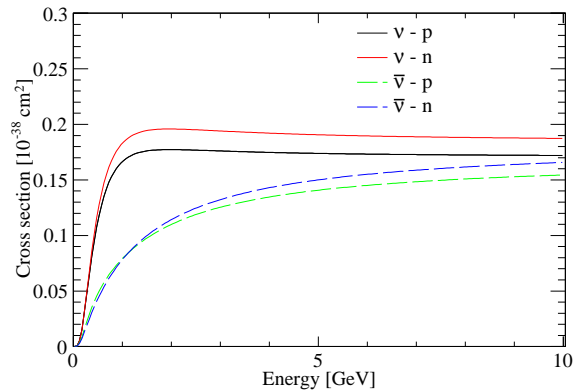}
	\caption{NCQE cross section on oxygen per nucleon as a function of neutrino energy. Solid lines show the neutrino cross sections with proton (black) and neutron (red), and dashed lines show antineutrinos cross section with proton (green) and neutron (blue).}
	\label{fig:TheoXsec}
\end{figure}

The NCQE cross section on oxygen in NEUT is simulated by an oxygen spectral function model~\cite{Benhar:2005dj} with the BBBA05 vector form factor~\cite{Bradford:2006yz} and the dipole parametrization of the axial form factor~\cite{Bradford:2006yz}, taking into consideration the Pauli blocking effect at $p_F=225$~MeV$/c$.
This cross section reproduces well the Ankowski model~\cite{Ankowski:2011ei} while providing additional information on kinematics of final state particles which cannot be directly extracted from the model.
The cross section per nucleon as a function of neutrino energy for neutrinos and antineutrinos is shown in Fig.~\ref{fig:TheoXsec}.

The interacting nucleus is left with a hole in $1p_{1/2}$, $1p_{3/2}$ or $1s_{1/2}$ state in the simple shell model.
The de-excitation $\gamma$ energy (and corresponding branching ratio) from the $1p_{3/2}$ state is 6.32 MeV (86.9\%) or 9.93 MeV (4.9\%) in the case of a proton knock-out, and 6.18 MeV (86.9\%) in the case of a neutron knock-out.
The spectroscopic factors follow the summary in Ref.~\cite{Abe:2014dyd}.
Other decay channels include further nucleon emission.
The branching ratios for the $1s_{1/2}$ hole state, which decays through several channels with gamma and nucleon emission, are based on the measurements of the $^{16}\text{O}(p,2p)^{15}\text{N}$ experiment RCNP-E148~\cite{Kobayashi:2006gb}.
Continuum states with multi-nucleon knock-out are assumed to have no gamma emission.
This assumption is taken into account in the estimation of systematic uncertainty on primary gamma emission.

\subsection{Detector simulation}

The SKDETSIM GEANT3 based~\cite{Brun:1987ma} simulation package is used to model particle propagation in the water, as well as the optical properties, photosensor and electronics response in SK.
This model has been tuned to match SK calibration data.

In interactions of atmospheric neutrinos with oxygen nuclei, hundreds of MeV are typically transferred to the struck nucleons, often resulting in the emission of further secondary nucleons and de-excitation gammas.
These secondary processes are simulated with the GCALOR package in GEANT3~\cite{Zeitnitz:1994bs}.
In particular, the NMTC~\cite{Bertini:1970zs} model is used for protons of all energies and neutrons above 20 MeV, while MICAP~\cite{osti_5567506} is used for neutrons below 20 MeV.
The discontinuity between the two models at 20 MeV is treated as a source of systematic uncertainty in secondary gamma emission.
There is no experimental data on the production of secondary particles in NCQE interactions in water.

\section{Event reconstruction and selection}
\label{sec:data}

\subsection{Data set}

This analysis uses SK-IV data collected in a fiducial volume 2 m away from the ID wall in all directions, containing 22.5 kton of water.
Due to a lowered threshold of SHE triggers since September 2011, the live-time is different for the kinetic energy range [9.5, 29.5] MeV and [7.5, 9.5] MeV, with 2,778 days and 1,886 days respectively.

This relatively high trigger threshold of 7.5 or 9.5 MeV is chosen to avoid the overlap with the dominant cosmic-ray-induced spallation background~\cite{Li:2015kpa}.
An upper limit of 29.5 MeV in electron kinetic energy is imposed to reduce backgrounds from atmospheric neutrino charged-current (CC) interactions, including Michel electrons~\cite{Bouchiat:1957zz} from decays of muons below the Cherenkov threshold.

Energy resolution and secondary gamma emission contribute to the reconstruction of NCQE events above the relatively high energy thresholds of 7.5 and 9.5 MeV.

\subsection{Data reduction}

One of the main motivations for the measurement of the NCQE cross section using atmospheric neutrinos is to determine the contamination of NC events in SRN searches.
Similar selection steps to those used in SRN searches~\cite{Zhang:2013tua} are used to select NCQE events.
These criteria can be categorized into the first reduction, the spallation cut, the further reductions, and the Cherenkov angle cut, the last of which distinguishes the data sample of SRN and NCQE events.
In addition, neutrons are tagged to further reduce the remaining backgrounds.
Neutron tagging is a technique to detect neutron capture on hydrogen, and will be covered in the next subsection.

\subsubsection{First reduction}

The following reduction cuts are first applied to remove spurious events and entering backgrounds~\cite{Abe:2016nxk, Abe:2017emr}.
First, a trigger cut is applied to remove calibration events and events that trigger the OD, which are most likely to be a cosmic-ray muon.
T2K beam events are also removed from the data sample by a beam trigger.
Then, a time difference cut is applied to remove events within 50 $\mu$s to the preceding cosmic-ray muon, which are likely to be decay-electrons or noise from energetic muons.
Afterwards, remaining low energy events are reconstructed with fitting goodness information.
If the fitting goodness is too low, the event is also excluded.
A fiducial volume cut is further applied to remove radioactivities from the PMTs.
The first reduction efficiency for the signal events is estimated to be $>99\%$.

\subsubsection{Spallation cut}

The spallation products from energetic muons traversing the detector form the main background in the 7.5-29.5 MeV energy range at SK~\cite{Li:2014sea, Li:2015lxa}.
This background increases rapidly at lower energy.

To remove spallation background while keeping signal efficiency, the time and track of a muon close to a low energy event are used.
A spallation-rich sample shortly after muon events and a random sample before muon events are used to construct a spallation likelihood function.
This function is the product of four parts depending on the transverse distance between the low energy vertex and the muon track, the longitudinal difference between the low energy vertex and the peak energy deposit of the muon, the amount of energy deposited near this peak, and the time difference between muon and candidate, respectively.
The maximum kinetic energy of spallation events is 20.6 MeV~\cite{Zhang:2015xra}. 
Considering energy resolution effects, spallation cuts are applied to all events with reconstructed energy up to 23.5 MeV.
If an event is highly likely to be a spallation event, it is removed from the data sample.
Spallation cuts are dependent on energy, with looser cuts for higher energy events.
Signal efficiency for these spallation cuts is estimated to be $90\%$ for the high energy bins above 17.5 MeV and $51\%$ for the energy bin at 7.5 MeV~\cite{Zhang:2015xra}.

\subsubsection{Further reduction}

External gamma rays from PMT surfaces and tank material may enter the fiducial volume and can be reconstructed as inward-pointing tracks.
We place a cut on the distance between the PMT surface and event vertex in the reconstructed direction of the candidate event~\cite{Bays:2011si}.
The efficiency for this cut is estimated to be $93\%$.

Other possible electron-like backgrounds include scattering of solar neutrinos off electrons and decay-electrons from unobserved muons.
Solar neutrinos form an important background in the energy range below 17.5 MeV.
Therefore, a solar angle cut is applied to remove solar neutrino events at $98\%$ efficiency~\cite{Abe:2016nxk}.
To account for the remaining background from atmospheric CC events, a pre-/post-activity cut which tags multiple peaks close in time, and a multi-ring cut which tags multiple Cherenkov rings close in space are applied~\cite{Bays:2011si, Zhang:2013tua}.
These cuts remove mis-reconstructed muons and the leakage of decay electron from these muons at high signal efficiencies of $95\%$ and $99\%$.
Muons and pions can also leak into the NC sample due to resolution smearing and mis-reconstructed Cherenkov angle~\cite{Zhang:2013tua}.
These events are tagged by the hit/charge pattern and the sharpness of their Cherenkov rings.

\subsubsection{Cherenkov angle cut}

For low energy analyses at SK, particle identification (PID) is performed using the opening angle of the Cherenkov ring in an event.
The Cherenkov angle is reconstructed using 3-hit combinations.
Given a reconstructed vertex, each set of 3-PMT hits uniquely defines a cone and its opening angle.
For each event, a histogram is filled with the opening angles for all 3-hit combinations in a TOF-subtracted time window of 15 ns and the bin with most entries is taken as the reconstructed Cherenkov angle for the event.

\begin{figure}[htbp] 
	\centering
	\includegraphics[width=0.89\columnwidth]{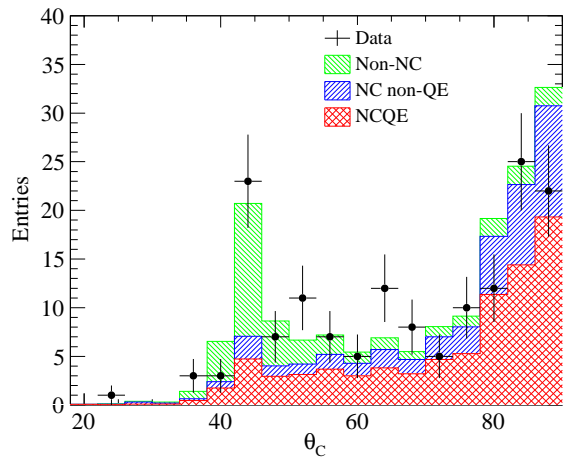}
	\caption{Cherenkov angle distribution in data (black points) and MC expectation (colored histograms) after all other cuts and with neutron tagging. The green histogram shows the non-NC backgrounds, blue shows the NC non-QE components, and red shows the NCQE components.}
	\label{fig:NC_thetaC}
\end{figure}

The Cherenkov angle of an electron above threshold peaks at $42^\circ$, while for a pion or muon below energy upper limit of 29.5 MeV, the opening angle is less than $42^\circ$.
Single $\gamma$ rays with the energy above the analysis threshold of 7.5 or 9.5 MeV are likely to produce multi-electrons during propagation, thus producing a more uniform Cherenkov light distribution.
The multi-$\gamma$'s from secondary processes further smear the Cherenkov light direction.
Therefore, the Cherenkov angle of NCQE events is usually reconstructed with high Cherenkov angles.
The angle distribution after neutron tagging is shown in Fig.~\ref{fig:NC_thetaC}.
Events with Cherenkov angles greater than $50^\circ$ are accepted as the signal at an efficiency of $86\%$, as summarized in Table~\ref{tab:RedEff}.

\begin{table}[h]
	\caption{Reduction efficiency and remaining sample size for each data reduction step without neutron tagging. Cuts are applied sequentially.}
	\begin{center}
	\begin{tabular}[c]{lcc} \hline\hline
		Reduction step & Signal eff. & Remaining events\\\hline
		First reduction & $>99\%$ & $\sim1,600,000$ \\
		Spallation cut & $77\%$ & $\sim170,000$ \\
		Incoming cut & $93\%$ & 73,348 \\
		Solar cut & $98\%$ & 64,037 \\
		Pre-/post-activity & $95\%$ & 56,650 \\
		Multi-ring & $99\%$ & 56,156 \\
		Cherenkov angle & $86\%$ & 27,577 \\\hline
		Total & $54\%$ & 27,577 \\\hline\hline
	\end{tabular}
	\label{tab:RedEff}
	\end{center}
\end{table}

\subsection{Neutron tagging}

NCQE events from atmospheric neutrinos are highly likely to knock out a nucleon from the oxygen nucleus.
Knocked-out neutrons will propagate in water and thermalize.
The thermalized neutron is then captured by a hydrogen nucleus and emits a single $\gamma$ at 2.2 MeV, which can be used to tag this event.

\subsubsection{Tagging algorithm}

The identification of neutrons emitted in IBD processes via captures on hydrogen in water Cherenkov detectors was initially developed for SRN searches~\cite{Zhang:2013tua}.
First, we calculate the hits within a sliding 10 ns TOF-subtracted time window and apply a pre-cut.
A hit cluster is selected as a candidate for neutron tagging only when the number of hits within this 10 ns time window exceeds 7.
Given the poor vertex resolution for 2.2 MeV gammas, the vertex of the prompt event is used in the TOF calculation.
Second, for every neutron candidate, a multilayer perceptron (MLP) is applied using derived parameters characterising hit time, hit pattern, and PMT charge.
A cut on the output of the MLP is used to select neutron events, and its efficiency is obtained using a sample where random trigger data is used to model the noise and the simulated signal events are superimposed~\cite{Zhang:2015xra}.

For this work, a recently upgraded neutron-tagging algorithm takes into consideration the reconstruction of neutron capture vertex using both the TOF-based algorithm and a brute-force fitter~\cite{Irvine:2014}.
It greatly enhances the discrimination power between neutron capture signals and accidental coincidence backgrounds.
The MLP cut value is optimized for every reconstructed prompt energy bin with the expected background and signal MC, imposing a more stringent cut for lower energy events, for which the background rate is higher.
This neutron-tagging technique has been validated with a calibration Am/Be source in SK-IV~\cite{Watanabe:2008ru}.

\subsubsection{Tagging NC neutrons}

Since the TOF-subtracted time is corrected with regards to the primary vertex, the distance between primary vertex and neutron capture vertex affects the neutron-tagging efficiency.
For IBD neutrons from SRN neutrinos, the neutron energy is sub-MeV and the neutron drifts typically less than 5\ cm away from the primary vertex.
However, for neutrons from NC events, their energy can reach several hundreds of MeVs and the neutrons can drift meters away; hence the neutron-tagging efficiency is dependent on neutron drift distance.
The difference in neutron-tagging efficiencies for different neutron energy profiles are shown in Table~\ref{tab:NtagEff}.
The average neutron-tagging efficiency for a single NCQE neutron in this analysis is 10.4\%.
The difference between SKDETSIM and another neutron propagation simulation in GEANT4~\cite{Agostinelli:2002hh} is taken as a source of systematic uncertainty.
\begin{table}[h]
	\caption{Neutron tagging efficiency for different neutron drift distances from different simulation setups with primary events at 14.5 MeV.
		Background here refers to accidental coincidence events.}
	\begin{center}
	\begin{tabular}[c]{lcccc} \hline\hline
		Process & Pre-cut & Pre-cut $+$ MLP\\\hline
		SRN & 35.4\% & 25.2\% \\ 
		NC & 29.8\% & 19.1\% \\
		Bkg. per event & 1.6 & 5.4$\times10^{-4}$ \\\hline\hline
	\end{tabular}
	\label{tab:NtagEff}
	\end{center}
\end{table}

\begin{figure}[!htbp] 
	\centering
	\includegraphics[width=0.89\columnwidth]{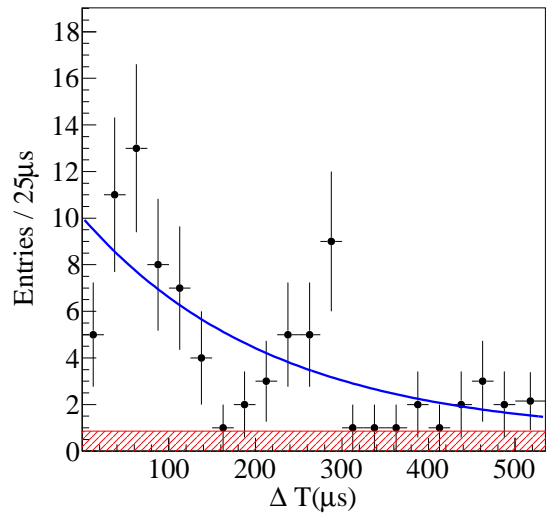}
	\caption{Neutron tagging time distribution for events with only one neutron tagged. The black points show the data distribution and the blue line shows the fitted exponential distribution. The red histogram indicates the accidental background.}
	\label{fig:Ntag_time}
\end{figure}

\begin{figure}[!htbp] 
	\centering
	\includegraphics[width=0.95\columnwidth]{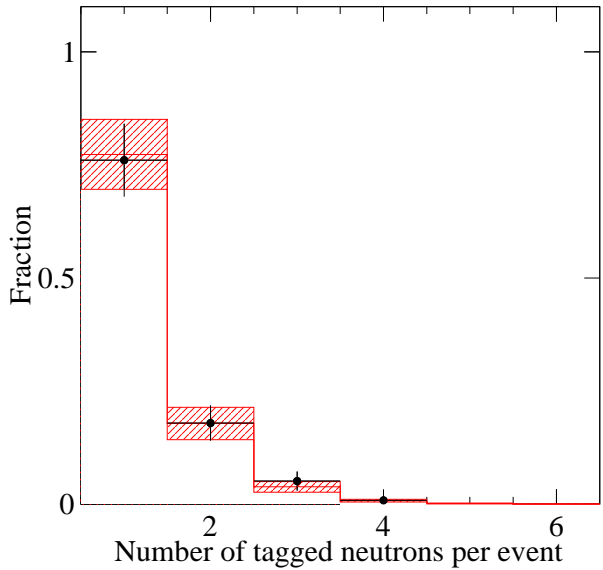}
	\caption{Neutron multiplicity in data (black points) and MC (red histogram). The uncertainties assigned to data are statistical while those assigned to MC are the 10\% intrinsic neutron-tagging uncertainty.}
	\label{fig:neutronmult}
\end{figure}

There are $117$ events with tagged neutrons remaining in the data sample, among which $89$ events have only one neutron tagged.
The neutron capture time for the NC sample after all reductions and with only one neutron is shown in Fig.~\ref{fig:Ntag_time}.
The accidental event rate can be estimated by using the same neutron-tagging criteria on a random trigger sample, and is calculated to be $13.7$ events in this NC sample.
Fitting to an exponential signal and a fixed background at the accidental background level, the total number of events with neutrons is $75.3\pm9.4$ and the life-time is given as $219.5\pm47.2\ \mu$s, consistent with the expectation of $204.7\ \mu$s~\cite{Watanabe:2008ru}.
Using a fixed $\tau$ at $204.7\ \mu$s fitting as a cross-check, the total number of events with neutrons is calculated to be $68.5\pm12.1$, consistent with fixed background fitting.
The neutron multiplicity comparison between data and MC is shown in Fig.~\ref{fig:neutronmult}.
The distributions are consistent within the uncertainties.

\begin{figure}[!htbp] 
	\centering
	\includegraphics[width=0.89\columnwidth]{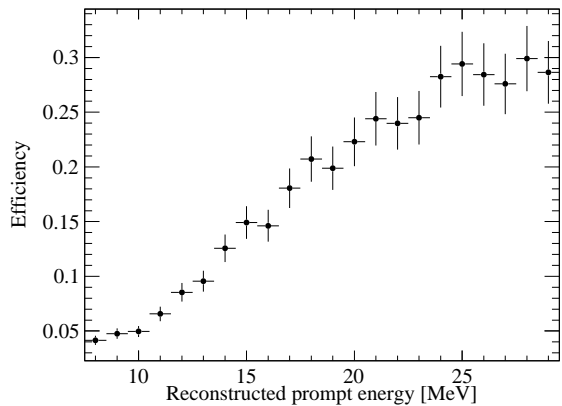}
	\caption{Reduction efficiency as a function of the reconstructed prompt energy.
	The energy dependence mainly comes from spallation cut (50\% to 100\%) and the neutron-tagging cut (4\% to 22\%).
	The error bars represent the reduction efficiency uncertainty and the neutron tagging uncertainty, as listed in Table.~\ref{tab:SysUncer}.
	}
	\label{fig:ReducEff}
\end{figure}

\begin{figure}[!htbp] 
	\centering
	\includegraphics[width=0.89\columnwidth]{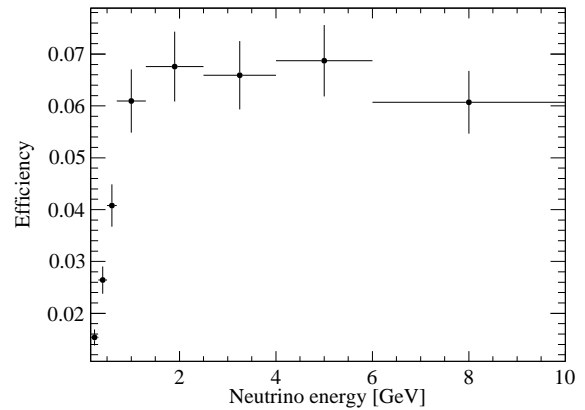}
	\caption{Detection efficiency as a function of the incident neutrino energy.
	The error bars represent the reduction efficiency uncertainty and the neutron tagging uncertainty.
	}
	\label{fig:ReducEffEnu}
\end{figure}

In this analysis, we select events with at least one neutron tagged.
The reduction efficiency including neutron tagging as a function of reconstructed prompt energy for NCQE gamma events within [7.5, 29.5] MeV is shown in Fig.~\ref{fig:ReducEff}.
This efficiency is relative to the NCQE events with neutron production and de-excitation $\gamma$ emission in the selected energy range.
The main efficiency sacrifice for events above the threshold comes from the spallation cut and neutron tagging.
These cuts are tuned per MeV energy bin towards a best significance on MC, putting an energy dependence on the final efficiency.
The overall detection efficiency relative to the number of true NCQE events in the fiducial volume with de-excitation $\gamma$'s is also calculated.
Figure~\ref{fig:ReducEffEnu} shows the overall detection efficiency as a function of the incident neutrino energy.
The higher detection efficiency for higher energy neutrinos originates from their higher neutron multiplicity.

\section{Cross-section extraction}
\label{sec:analysis}

The $117$ events in the final data sample include NCQE, NC non-QE channels, neutron backgrounds including leakage from reactor neutrinos and spallation, and accidental backgrounds.
To evaluate the contribution from non-NC and NC non-QE events, and to derive the NCQE cross section, further analysis is performed using MC and different data samples.

\subsection{Observed events}

\begin{figure}[htbp] 
	\centering
	\includegraphics[width=0.89\columnwidth]{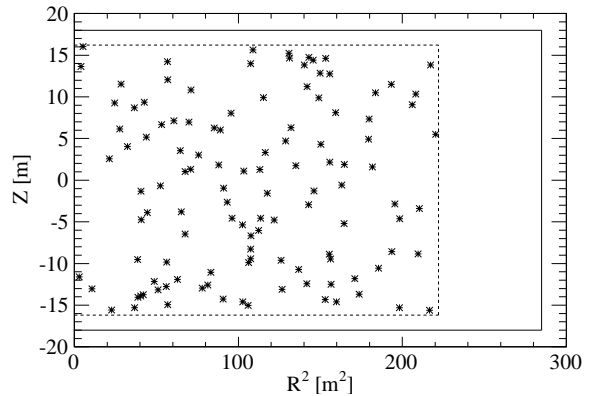}
	\caption{Vertex distributions of the NC sample. The solid rectangle shows the inner detector of SK, and the dotted rectangle shows the fiducial volume region.}
	\label{fig:NC_basic}
\end{figure}

\begin{figure}[!htbp] 
	\centering
	\includegraphics[width=0.89\columnwidth]{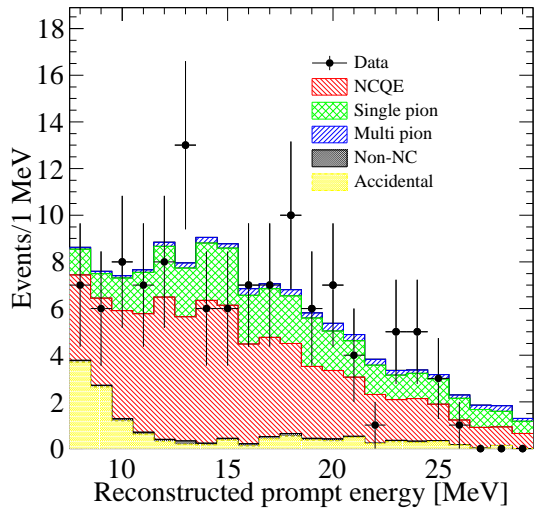}
	\caption{Energy distribution of the NC data sample (black points) against MC expectations (colored histograms).
	The red histogram shows NCQE signal, green shows single-pion channels, blue shows multi-pion channels, black shows non-NC backgrounds, and yellow shows the accidental background without any true neutron. Summing up the green and the blue histogram gives the NC non-QE backgrounds.}
	\label{fig:NC_dataMC}
\end{figure}

The vertex distribution of the final NC sample is shown in Fig.~\ref{fig:NC_basic}.
The vertices are uniformly distributed and consistent with the expectations of the NC signal.
The energy distribution is shown Fig.~\ref{fig:NC_dataMC}.
Note that the energy dependent efficiency shown in Fig.~\ref{fig:ReducEff} is not corrected in this figure.

Besides accidental background, non-NC backgrounds in the data sample include leakage from spallations, reactor neutrinos, as well as atmospheric neutrino CC interactions.
The spallation background is estimated using a spallation data sample derived by the local proximity to a muon in both time and space~\cite{Zhang:2015xra}.
The same cuts as in the NC event selection are then applied to this data sample, and the number of spallation background is estimated to be $0.5$, mainly contributed by $^9$Li.
Reactor neutrinos produce IBD positrons, for which the event rate is predicted by a simulation based on the IAEA data~\cite{IAEA:2017}.
Their leakage into the NC sample through Cherenkov angle resolution is estimated to be $0.1$ event.
Atmospheric neutrino CC backgrounds include $\nu_e$ CC and $\nu_\mu$ CC.
This background is estimated from MC to be 0.4 events for $\nu_e$ CC and 0.8 events for $\nu_\mu$ CC.
These non-NC backgrounds are summarized in Table.~\ref{tab:datacompo} and compared with the MC predictions in NCQE and other NC channels.
\begin{table}[h]
	\caption{Predictions of components in the final data sample and the comparison with signal MC.}
	\begin{center}
	\begin{tabular}[c]{lc} \hline\hline
		Components & Events \\\hline
		NC single $\pi$ & 33.4 \\
		NC multi-$\pi$'s & 4.2 \\
		DIS & 0.0 \\
		$\bar\nu_e$ CC & 0.4 \\
		$\bar\nu_\mu$ CC & 0.8 \\
		Accidental & 13.7 \\
		Spallation & 0.5 \\
		Reactor & 0.1 \\\hline
		Total background & 53.1 \\
		Observed data & 117 \\\hline
		Background subtracted data & 63.9 \\\hline
		NEUT NCQE prediction & 71.9 \\\hline\hline
	\end{tabular}
	\label{tab:datacompo}
	\end{center}
\end{table}

\subsection{Measured cross section}

The NCQE cross section is measured by comparing data and MC expectation.
The theoretical prediction of flux-averaged NCQE cross section on oxygen, $<\sigma^\text{theory}_\text{NCQE}>$, can be expressed as:
\begin{equation}
	\begin{aligned}
		<\sigma^\text{theory}_\text{NCQE}>&=\frac{\int_\text{160 MeV}^\text{10 GeV}\sum\limits_{i=\nu,\bar\nu}\phi_i(E_\nu)\times\sigma_i(E_\nu)^\text{theory}_\text{NCQE}dE_\nu}{\int_\text{160 MeV}^\text{10 GeV}\sum\limits_{i=\nu,\bar\nu}\phi_i(E_\nu)dE_\nu}\\
		&=1.14\times10^{-38}\text{ cm}^2,
	\end{aligned}
\end{equation}
where $\phi_i(E_\nu)$ is the atmospheric neutrino flux at neutrino energy $E_\nu$, $\sigma_i(E_\nu)$ is the corresponding theoretical cross section, and $i$ sums for neutrino and antineutrino species.
The energy integral is performed between 160 MeV and 10 GeV, above which the atmospheric neutrino flux is decreasing rapidly, and below which the NCQE cross section is very small.
The uncertainty introduced by the cutoff is evaluated and included as a systematic error.
The measured cross section is therefore expressed as:
\begin{equation}
	\begin{aligned}
		<\sigma^\text{observed}_\text{NCQE}>=&\frac{N_\text{tot}^\text{obs}-N_\text{acc}^\text{exp}-N_\text{others}^\text{exp}-N_\text{NCothers}^\text{exp}}{N_\text{NCQE}^\text{exp}}\\
		&\times<\sigma^\text{theory}_\text{NCQE}>\\
		=&(1.01\pm0.17_\text{stat.})\times10^{-38}\text{ cm}^2,
	\end{aligned}
\end{equation}
where $N_\text{tot}^\text{obs}$ refers to the observed number of events in the final reduction sample, $N_\text{acc}^\text{exp}$ refers to the accidental background without a true neutron, which is evaluated from random trigger data, and $N_\text{others}^\text{exp}$ refers to the non-NC backgrounds including spallation products, reactor antineutrinos, and atmospheric neutrino CC interactions.
The term $N_\text{NCothers}^\text{exp}$ refers to the expected number of NC non-QE backgrounds, and $N_\text{NCQE}^\text{exp}$ refers to the expected number of NCQE events.

\subsection{Systematic uncertainties}

Atmospheric neutrino flux uncertainty varies for different energy bins, as given in Ref.~\cite{Richard:2015aua}.
In this paper, a conservative estimation at $18\%$ in [160 MeV, 10 GeV] is taken.
Atmospheric $\nu/\bar\nu$ ratio has $5\%$ uncertainty~\cite{Honda:2006qj}.
The cross sections for NC processes other than NCQE have $18\%$ uncertainty~\cite{Abe:2014dyd}.
The spectroscopic factors and the gamma emission branching ratios determines the uncertainty from primary simulation~\cite{Huang:2016}.
The uncertainty from secondary gamma emission is estimated by varying the neutron energy profile and neutron multiplicity from the simulation~\cite{Huang:2016}.

Since this analysis requires at least one neutron to be detected via neutron tagging on hydrogen, an additional systematic uncertainty arises from the uncertainty of neutron multiplicity in neutral-current interactions.
There are no data available to estimate neutron multiplicity in NCQE process for oxygen, so this uncertainty is evaluated by comparing different simulations.
We compared the multiplicity and spectra of primary neutrons predicted by NEUT and GENIE~\cite{Andreopoulos:2009rq}, and the secondary neutron production as well as neutron thermalization and capture predicted by SKDETSIM and GEANT4 for different energy spectra of the primary neutrons.
Taking the result from NEUT and SKDETSIM as the central value, the difference in neutron multiplicity predictions after applying the tagging efficiency is $12\%$ for NCQE events.
For a conservative estimation, we take the difference in neutron multiplicity prediction at 100\% detection effciency, $21\%$, as the systematic uncertainty.
The neutron energy spectra from GENIE introduces $18\%$ deviation the NEUT spectra for NCQE events and $14\%$ for NC non-QE events.
The GEANT4 simulation of neutron transportation introduces $+7\%$ deviation from SKDETSIM for NCQE events, and $+4\%$ for NC non-QE events.
We assign asymmetric uncertainty to the neutron transportation simulation term, and leave the neutron multiplicity and energy spectra uncertainty to be symmetric.

Data reduction besides neutron tagging imposes a $3\%$ systematic uncertainty.
Neutron tagging efficiency has $10\%$ intrinsic uncertainty from calibration (Am/Be) and MC for low-energy neutrons.
The cutoff at 10 GeV imposes a $0.1\%$ uncertainty using simulation with the measured high energy atmospheric neutrino flux~\cite{Richard:2015aua}.
The cutoff at 160 MeV imposes $<0.7\%$ uncertainty, which is estimated by simulation with the theoretical prediction of low energy atmospheric neutrino flux~\cite{Battistoni:2005pd}.
The evaluation of non-NC (reactor, $^9$Li, CC, etc) leakage into NC sample imposes $21\%$ uncertainties to $N_\text{others}^\text{exp}$, but due to the small ratio of events from non-NC background, this uncertainty propagates to only $0.2\%$ on the final cross section result.

\begin{figure}[htbp] 
	\centering
	\includegraphics[width=0.89\columnwidth]{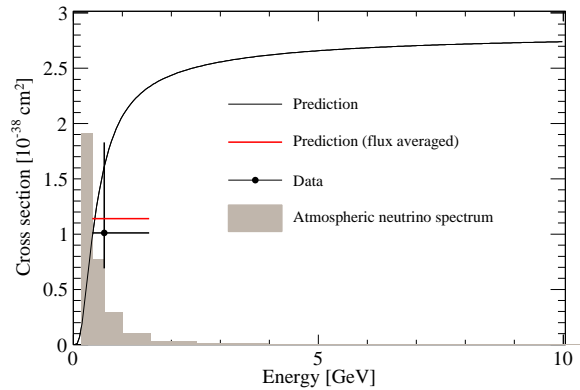}
	\caption{The gray histogram shows the atmospheric neutrino spectrum predicted by HKKM model, the black curve shows the cross section from Ankowski model, the red line shows the predicted flux-averaged cross section, and the black point shows the measured cross section.}
	\label{fig:Xsec}
\end{figure}

All the uncertainties are listed in Table.~\ref{tab:SysUncer}.
To account for the correlations including the flux uncertainty and the reduction uncertainty between the NCQE sample and other samples, A toy-MC is used to derive the uncertainty envelope for the NCQE cross section.
The 68\% confidence level region is finally calculated as $[0.69, 1.83]\times10^{-38}$ cm$^2$, and the cross section is measured to be $(1.01\pm0.17(\text{stat.})^{+0.78}_{-0.30}(\text{sys.}))\times10^{-38}$ cm$^2$, as shown in Fig.~\ref{fig:Xsec}.

\begin{table}[h]
	\caption{Uncertainties in NCQE measurement}
	\begin{center}
	\begin{tabular}{l c c}\hline\hline
		& NCQE & NC non-QE\\\hline
		$\nu_\text{atm}$ flux & \multicolumn{2}{c}{18\%} \\
		$\nu/\bar\nu$ ratio & \multicolumn{2}{c}{5\%} \\\hline
		Cross-section & & 18\%\\
		Primary $\gamma$'s & 15\% & 3\%\\
		Secondary $\gamma$'s & 13\% & 13\%\\
		Neutron multiplicity & 21\% & 16\%\\
		Neutron energy & 18\% & 14\%\\
		Neutron transportation & $+7\%$ & $+4\%$\\
		\hline
		Data reduction & \multicolumn{2}{c}{3\%} \\
		Neutron tagging & \multicolumn{2}{c}{10\%} \\
		Others & \multicolumn{2}{c}{$0.7\%$}\\\hline\hline
	\end{tabular}
	\label{tab:SysUncer}
	\end{center}
\end{table}

\subsection{Discussion on future improvement}

The uncertainty in this measurement is dominated by systematic uncertainties including the atmospheric flux, cross section of other NC processes, primary and secondary process simulation, neutron simulation, as well as neutron-tagging efficiency.
The flux measurement will improve with future Cherenkov detectors such as Hyper-Kamiokande~\cite{Abe:2011ts}. 
The cross section for other NC processes can be improved by the T2K off-axis near detector ND280~\cite{Abe:2013jth, Abe:2016aoo} and other experiments such as MiniBooNE~\cite{AguilarArevalo:2010bm} and MINERvA~\cite{McGivern:2016bwh}.
Hadron production experiments such as EMPHATIC~\cite{Gameil_2018} will also contribute to reducing flux uncertanties.
For the simulation of primary and secondary processes, the gamma ray emission experiment at RCNP is likely to reduce the uncertainty soon~\cite{Ashida:2018vsn, Ou:2016pkv}.

The statistics in this analysis is limited by the neutron-tagging efficiency and the energy threshold.
The present efficiency for NCQE neutrons in pure water is relatively poor at $4-22\%$.
When SK updates to SK-Gd~\cite{Beacom:2003nk, Sekiya:2017lgj}, the efficiency would increase to about 80\% due to the higher total energy of the $\gamma$ cascades.
A measurement of neutron multiplicity will also provide contraints on the simulation of neutron production.
Besides, at SK-Gd, the neutron capture signal can trigger the detector directly, and thus the lower energy threshold of this analysis for prompt $\gamma$'s will not be limited by the SHE trigger threshold.
Lowering the analysis threshold to 3.5 MeV will double the detection efficiency of NCQE $\gamma$ events.

\section{Conclusion}
\label{sec:conclusion}

The first measurement of the NCQE cross section with atmospheric neutrinos on oxygen is reported.
NCQE events are selected by the nuclear de-excitation gamma and neutron capture signal on hydrogen.
The neutron-tagging technique is employed to enhance the signal-background separation.
We obtained $117$ events after data reduction, in agreement with the expectation of $125.0$, including $71.9$ estimated from NCQE channel, $37.6$ from non-QE NC channels, and $15.5$ from non-NC background.
The NCQE cross section averaged over the atmospheric neutrino flux at SK is measured to be $(1.01\pm0.17(\text{stat.})^{+0.78}_{-0.30}(\text{sys.}))\times10^{-38}$ cm$^2$, consistent with the theoretical prediction of $1.14\times10^{-38}\text{ cm}^2$.

This result improves the estimation of NCQE component in low energy rare signal detection in water Cherenkov detectors, especially in the search of SRNs.
It will also benefit future water Cherenkov experiments, such as SK-Gd and Hyper-K.

\section*{Acknowledgments}
We gratefully acknowledge the cooperation of the Kamioka Mining and Smelting Company.
The Super‐Kamiokande experiment has been built and operated from funding by the Japanese Ministry of Education, Culture, Sports, Science and Technology, the U.S. Department of Energy, and the U.S. National Science Foundation.
Some of us have been supported by funds from the National Research Foundation of Korea NRF‐2009‐0083526 (KNRC) funded by the Ministry of Science, ICT, and Future Planning, the European Union H2020 RISE‐GA641540‐SKPLUS, the Japan Society for the Promotion of Science, the National Natural Science Foundation of China under Grants No. 11620101004, the National Science and Engineering Research Council (NSERC) of Canada, the Scinet and Westgrid consortia of Compute Canada, and the National Science Centre, Poland (2015/17/N/ST2/04064,2015/18/E/ST2/00758).

\nocite{*}

\bibliography{apssamp}

\end{document}